\documentclass{INTERSPEECH2023}
\usepackage{multirow}

\interspeechcameraready

\title{Stochastic Pitch Prediction Improves the Diversity and Naturalness of Speech in Glow-TTS}
\name{Sewade Ogun, Vincent Colotte, Emmanuel Vincent}
\address{Université de Lorraine, CNRS, Inria, LORIA, F-54000 Nancy, France}
\email{sewade.ogun@inria.fr, vincent.colotte@loria.fr, emmanuel.vincent@inria.fr}

\begin{document}

\maketitle
 
\begin{abstract}

Flow-based generative models are widely used in text-to-speech (TTS) systems to learn the distribution of audio features (e.g., Mel-spectrograms) given the input tokens and to sample from this distribution to generate diverse utterances. However, in the zero-shot multi-speaker TTS scenario, the generated utterances lack diversity and naturalness.
In this paper, we propose to improve the diversity of utterances by explicitly learning the distribution of fundamental frequency sequences (pitch contours) of each speaker during training using a stochastic flow-based pitch predictor, then conditioning the model on generated pitch contours during inference.
The experimental results demonstrate that the proposed method yields a significant improvement in the naturalness and diversity of speech generated by a Glow-TTS model that uses explicit stochastic pitch prediction, over a Glow-TTS baseline and an improved Glow-TTS model that uses a stochastic duration predictor.

\end{abstract}

\section{Introduction}

Multi-speaker text-to-speech (TTS) systems have significantly improved in quality and naturalness over the years. Given the one-to-many mapping between phonemes and a speaker's utterance, and other variabilities which improve the perceived quality of speech, some systems model these variabilities by explicitly learning the speaker's characteristics such as pitch, energy, or rhythm \cite{renfastspeech, lancucki2021fastpitch}. Instead, flow-based TTS systems learn the distribution of the Mel-spectrogram (or audio), which includes all the speaker's characteristics, given the input tokens and a speaker embedding \cite{kim2020glow, valleflowtron, kim2021conditional, miao2020flow}. Flow-based generative models are able to fit complex data distributions and generate diverse utterances with high quality. In addition, they eliminate the problem of over-smoothness in generated Mel-spectrograms \cite{ren2022revisiting}, and therefore, they do not require a post-processing stage for generated Mel-spectrograms, as is found in many non-flow-based non-generative models, e.g., \cite{shen2018natural}.

However, in the zero-shot multi-speaker TTS scenario, flow-based models are unable to learn the diversity of speaking styles inherent to each speaker given only the speaker embeddings. In particular, when these models are used to generate speech for unseen speakers, the generated utterances often sound monotonic and less natural. Our goal is to design an improved flow-based architecture that is able to match the diversity of real speech, including for unseen speakers.

Many methods have been proposed in the literature for controlling the expressivity of speakers. These methods use global style tokens or embeddings learned during training \cite{wang2018style, valle2020mellotron, min2021meta} to control the speaking style. Style tags \cite{kim21n_interspeech} have also been used to transfer speaking styles from text embeddings to speech utterances, while other studies, e.g., \cite{9616249}, have attempted to transfer emotions from one speaker to another using an expressivity encoder. These methods generally aim to mimic the style of a reference utterance or text, which is different from our goal.

Several TTS models also explicitly model speech properties such as duration, energy, timbre, rhythm, or speaking rate. In FastSpeech2 \cite{renfastspeech} and FastPitch \cite{lancucki2021fastpitch}, the duration, pitch contour, and energy of the input tokens are explicitly learned. 
ZSM-SS \cite{kumar21c_interspeech} uses a similar approach with external speaker embeddings, but it normalizes the prosodic features (e.g., energy and pitch) in the affine layers of the model to improve speaker similarity in the zero-shot multi-speaker setting. Since these systems are deterministic by design, they are unable to output diverse renderings for a given text input and speaker.
Recently, Varianceflow \cite{lee2022varianceflow} replaced the deterministic pitch and energy predictors in a FastSpeech2 TTS model with normalizing flows. The work showed the effectiveness of using normalizing flows to learn the prosodic features, however, it differs from ours as it was proposed for a single-speaker, non-generative TTS model.

SC-GlowTTS \cite{casanova21b_interspeech} is a zero-shot multi-speaker TTS system that conditions a Glow-TTS model on external speaker embeddings. SNAC \cite{choi2022snac}, also a flow-based zero-shot multi-speaker TTS system, uses a coupling layer that explicitly normalizes the input by the parameters predicted from a speaker embedding vector. VITS \cite{kim2021conditional}, an end-to-end flow-based TTS model, introduced a stochastic flow-based duration predictor to generate diverse rhythms from the input text, which was found to improve the rhythm of generated utterances over a deterministic duration predictor. Your-TTS extends VITS to the multi-speaker multi-language scenario, enabling the generation of utterances for unseen speakers in several languages. Although all these models inherently enable sampling from the latent representation of the input tokens to generate diverse utterances, the quality of generated utterances in the multi-speaker setup is still worse than that of single-speaker TTS models \cite{ren2022revisiting}.

The focus of this work is on generating realistic and diverse utterances for speakers unseen during training by explicitly learning the pitch distribution. To do so, we enhance a Glow-TTS system by integrating a stochastic duration predictor and a stochastic pitch predictor into the model. The stochastic duration predictor learns more realistic phoneme duration and gives better speech rhythm to the utterances while the pitch predictor learns the distribution of pitch contours in the dataset. These provide a method to generate diverse and natural utterances for different unseen speakers at inference time.

Section~\ref{section:method} presents the proposed improved Glow-TTS architecture. Sections~\ref{section:experiment} and \ref{section:result} describe the experimental setup and the results. We conclude in Section~\ref{section:conclusion}.

\section{Improved Glow-TTS Architecture}
\label{section:method}
The classical Glow-TTS architecture comprises a Transformer-based encoder, a flow-based decoder, and a deterministic duration predictor. The Transformer encoder generates contextual phonetic embeddings from the input tokens, which are then linearly projected into an 80-dimensional representation of the mean $\mu$ of the prior distribution. 
Given $\mu$, a scalar noise temperature $T$, and a random vector $\epsilon$ sampled from the standard normal distribution, the latent representation $z$ sampled from the prior distribution can be expressed as
\begin{align}
  z &= \mu + T\,\epsilon.
  \label{eq:1}
\end{align}
At training time, only the mean $\mu$ is predicted and the temperature $T$ is fixed to $1$. At inference time, a certain value of $T$ (often smaller than $1$) is chosen, and a latent representation is sampled from the prior distribution as input to the decoder to generate a Mel-spectrogram.

\begin{figure}[h!]
  \centering
  \includegraphics[width=7.6cm,trim=0 10 0 0,clip]{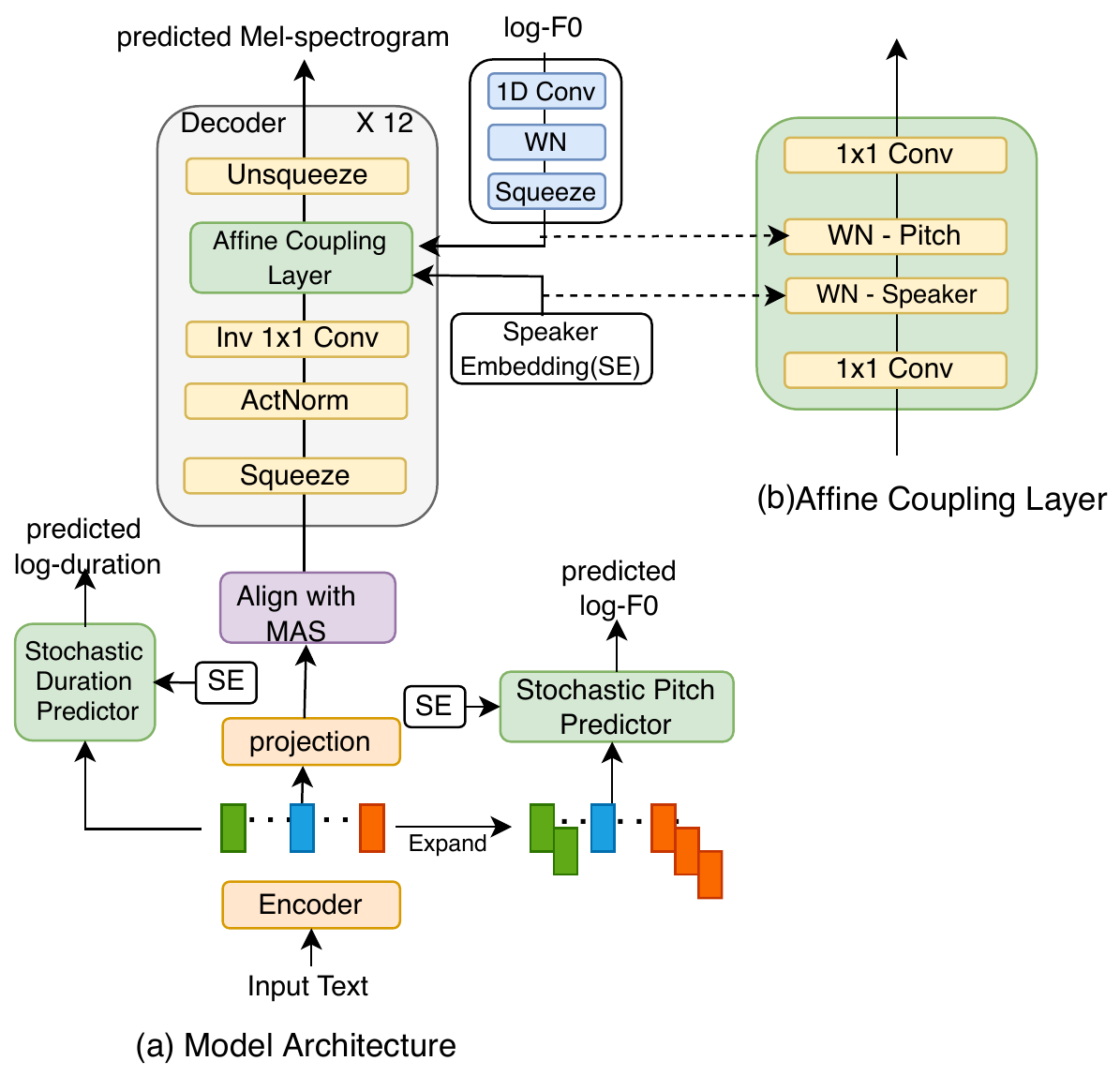}
  \caption{Proposed Improved Glow-TTS architecture including (a) a Stochastic Duration Predictor and a Stochastic Pitch Predictor, and (b) an Affine coupling layer with pitch and speaker conditioning. WN represents Weight Normalization layers.}
  \label{fig:stochastic_pitch_pred}
\end{figure}

Figure \ref{fig:stochastic_pitch_pred} illustrates the proposed improved architecture with arrows showing the computation during inference. This architecture incorporates a stochastic duration predictor and a novel stochastic pitch predictor, and it requires changes to the decoder.

\subsection{Decoder}
\label{section:decoder}
The decoder is a stack of normalizing flow blocks, each consisting of an activation normalization layer, an invertible 1x1 convolution layer, and an affine coupling layer. Classically, in the multi-speaker TTS setup, the decoder is conditioned on speaker embeddings through the affine coupling layers. At training time, the decoder inverts the Mel-spectrogram into a hidden representation, which is aligned to the latent representation from the encoder using an alignment search algorithm called Monotonic Alignment Search (MAS). MAS finds the most probable monotonic alignment between the hidden representation and the mean of the prior distribution. 

In this work, the decoder is additionally conditioned on the pitch information. We use the log-F0 values extracted from the ground-truth Mel-spectrograms as pitch targets, and set all unvoiced values to zero during training. The log-F0 is first projected to match the dimension of the decoder features using a 1D convolution layer, which is then used to condition the decoder features through the affine coupling layers of the decoder. We squeeze the projected log-F0 features to match the squeeze ratio of the original decoder features. At inference, the predicted log-F0 values generated by the stochastic pitch predictor are used for conditioning.

\subsection{Duration Prediction}


The classical Glow-TTS model predicts the speaker-dependent duration of each token through conditioning on speaker embeddings. The predicted durations have been found to yield unnatural rhythm and cannot express how a person speaks at different speaking rates \cite{kim2021conditional}.

To generate a more human-like rhythm, we replace the deterministic predictor by the stochastic flow-based duration predictor\footnote{We used the official implementation at \url{https://github.com/jaywalnut310/vits}.} in VITS \cite{kim2021conditional}. This model is composed of two normalizing flow stages that learn the discrete duration estimated using the MAS alignment algorithm. At training time, a first-stage set of flows (called post flows) inverts the discrete duration which is conditioned on the speaker and input embeddings into a de-discretized distribution of log-duration, then the second-stage set of flows inverts these log-durations into a Gaussian noise.\footnote{\label{stopgradient}A stop-gradient operator is applied to the input embeddings when learning the duration and pitch.} At inference time, the post-flows are discarded. A Gaussian noise vector is simply sampled and passed through the second-stage flows to generate log-durations.

\subsection{Pitch Prediction}

Utterances with diverse stress and intonation can be generated by sampling different $\epsilon$ in \eqref{eq:1}.
In the multi-speaker TTS scenario, the resulting distribution of stress and intonation varies depending on the speaker embedding.
We found that this alone is not enough to produce diverse, natural, and lively utterances for speakers unseen in training.

Thus, we use an explicit flow-based stochastic pitch predictor to learn pitch contours that closely match the distribution of real contours.
This model is made up of 4 blocks of 1D convolution, dilated and depth-separable convolutions and spline flows \cite{durkan2019neural}. Variational data augmentation \cite{chen2020vflow} is applied to increase the dimension of the scalar pitch inputs, as this is necessary to perform high-dimensional transformation. Specifically, an extra vector of Gaussian noise is concatenated to the pitch vector or input Gaussian noise as padding, before passing them through the block of flows. Our architecture is borrowed from the second set of flow blocks of the stochastic duration predictor.

During training, the speaker-conditioned input embeddings are expanded to match the duration of each input token estimated by MAS. These are then used as input to the flow-based stochastic pitch predictor to reverse the log-F0 values into Gaussian noise. The model is optimized jointly with other modules via maximum likelihood estimation.\footref{stopgradient} At inference, the speaker-conditioned input embeddings are expanded using the predicted duration values, while a Gaussian noise is sampled to generate log-F0 values for the decoder, as shown in Fig.~\ref{fig:stochastic_pitch_pred}.

\section{Experimental Setup}
\label{section:experiment}
We conducted objective and subjective evaluations on three variants of the model: the baseline Glow-TTS (54.5~M parameters), Glow-TTS with stochastic duration predictor (GlowTTS-STD, 55.3~M parameters), and Glow-TTS with stochastic duration and pitch predictors (GlowTTS-STDP, 56.0~M parameters).

\subsection{Dataset Preparation}

The Common Voice dataset (English subset, version 7.0) \cite{ardila2020common} was used to train all the models. Common Voice is a crowdsourced, read speech dataset with very large speaker diversity. The stochastic pitch predictor can benefit from learning a diverse distribution of F0 values for speakers in different contexts from the dataset. The dataset was filtered by selecting only high-quality speakers based on speaker-level Mean Opinion Scores (MOS) estimated by a MOS estimator \cite{ogun2023can}. We used the resulting filtered 4,469-speaker, 230.75 h dataset for model training and validation. 512 samples were randomly selected from the dataset for validation. The dataset was resampled from 48~kHz or 32~kHz to 16~kHz, and silences and long pauses were removed using a voice activity detection tool.\footnote{\url{https://github.com/wiseman/py-webrtcvad}}

F0 contours were extracted from the training and validation sets using pYIN \cite{mauch2014pyin} with 64~ms window size and 16~ms hop size to match the parameters of the Mel-spectrogram. We found it suitable in order to aid the correct alignment of the extracted F0 values to the input embeddings during model training. In addition, a speaker embedding was pre-computed for each utterance from a speaker verification model\footnote{\url{https://github.com/resemble-ai/Resemblyzer}} trained on Voxceleb \cite{Nagrani19}. 
The embeddings are l2-normalized, 256-dimensional vectors. Phonemes (without stress) and characters were used as input tokens in a mixed approach \cite{kastner2019representation}. The input tokens were interspersed with a blank token.

\subsection{Training and Inference Hyperparameters}
All TTS models were trained in automatic mixed-precision mode using a global batch size of 192 on 4 Nvidia RTX GPUs. Optimization was done using the RAdam optimizer \cite{liuvariance} with a learning rate of 0.001, and a cosine-annealing scheduler with a linear warm-up of 6,000 steps. It took approximately 4 days for the models with stochastic pitch prediction and stochastic duration prediction to completely train while the baseline Glow-TTS was trained for 2 days. Each model was trained for 200 epochs, and the best model was selected at the end of the training run using the validation loss.

At inference time, the noise temperature of the prior distribution was set to 0.667 for all models as in Glow-TTS \cite{kim2020glow}. The noise temperature for the pitch predictor was set to 0.8, similar to VITS \cite{kim2021conditional}. The noise temperature for the duration predictor was also set to 0.8 for speaking style evaluation and to 1.0 for reading style evaluation (see Section~\ref{subsec:subjective}).

A 16~kHz HiFi-GAN V1 vocoder \cite{kong2020hifi} trained on the LibriTTS dataset \cite{zen2019libritts} was used to convert the generated Mel-spectrograms into audio. The vocoder was not finetuned on Mel-spectrograms generated by the models. All experiments were performed using the NeMo toolkit \cite{kuchaiev2019nemo}.

\subsection{Subjective Evaluation}
\label{subsec:subjective}
For all the subjective evaluations, 3 unseen male (p245, p254, and p263) and 3 unseen female speakers (p228, p231, and p250) were selected from the VCTK dataset \cite{veaux2017cstr}.\footnote{\label{foot:embed} Speaker embeddings in all of our evaluations were extracted from the fifth utterance (SpeakerID 005) of the VCTK speaker.}
Firstly, the naturalness of the utterances generated by each model both in speaking style \textbf{(N-MOS)} and reading style \textbf{(NR-MOS)} were evaluated subjectively. We selected 3 short (3--4~s) text sentences from VCTK for the N-MOS evaluation, and 3 long (9--13~s) text sentences with punctuation from LibriTTS for the NR-MOS evaluation to generate 18 short and 18 long utterances per model. The evaluators were asked to rate how likely these are real speaking and read speech, respectively. Secondly, the speaker similarity \textbf{(S-MOS)} of the generated utterances to a reference utterance was evaluated. For this, 3 utterances that were recorded by each of the selected speakers in the VCTK dataset were used. The VCTK reference utterances were re-synthesized using the pretrained vocoder (VCTK-copy) and were added to the subjective evaluations to determine the upper bound on the scores. Lastly, the diversity of generated utterances \textbf{(D-MOS)} was evaluated for each model. For each speaker, 3 utterances were generated from the same text prompt and concatenated. The evaluators were asked to rate the diversity of the utterances in terms of the intonation (flow of pitch) of the speaker.
Mean opinion score evaluations were performed on a 1--5 Likert scale. In total, 26 volunteers participated in the evaluation.

\begin{table*}[ht!]
\centering
\caption{N-MOS, NR-MOS, S-MOS of utterances generated for 6 unseen speakers by the baseline Glow-TTS, GlowTTS-STD, GlowTTS-STDP, and VCTK-copy. Bold numbers denote the best system in each row and the systems statistically equivalent to it.} 
 \label{tab:mos_table_1}
\begin{tabular}{|ll|c|c|c||c|}
\cline{3-6}
\multicolumn{2}{l|}{}  & Baseline        & GlowTTS-STD        & GlowTTS-STDP  & VCTK-copy        \\ \hline
\multirow{3}{*}{N-MOS}  & Male     & $2.92$       & $3.11$       & $\mathbf{3.40}$               & $4.00$               \\
                      & Female     & $3.35$       & $\mathbf{3.51}$       & $\mathbf{3.51}$             & $4.40$               \\
                      & Total & $3.13$       & $3.31$       & $\mathbf{3.45}$               & $4.21$               \\ \hline 
\multicolumn{2}{|l|}{WV-MOS}      & $4.11$       & $\mathbf{4.17}$       & $\mathbf{4.18}$         & $4.32$              \\ \hline\hline
\multirow{3}{*}{NR-MOS} & Male     & $2.91$       & $\mathbf{3.24}$       & $\mathbf{3.25}$       & -       \\
                      & Female     & $2.98$       & $3.28$       & $\mathbf{3.53}$               & -       \\
                      & Total & $2.95$       & $3.26$       & $\mathbf{3.39}$               & -       \\ \hline\hline
\multirow{3}{*}{S-MOS} & Male     & $2.43$       & $2.90$       & $\mathbf{3.10}$               & $4.71$       \\
                      & Female     & $2.14$       & $\mathbf{3.07}$       & $2.91$               & $4.85$       \\
                      & Total & $2.26$       & $\mathbf{2.98}$       & $\mathbf{2.99}$               & $4.79$       \\ \hline
\multicolumn{2}{|l|}{cos-sim}      & $0.8287$       & $\mathbf{0.8364}$       & $\mathbf{0.8319}$         &  $0.8121$              \\ \hline
\end{tabular}
\end{table*}

\subsection{Objective Evaluation}
Objective evaluations were carried out similarly to subjective evaluations. The overall quality \textbf{(WV-MOS)} of the utterances generated by each model was evaluated using a MOS predictor \cite[App.~C]{andreev2022hifi++}.\footnote{pretrained model can be found at \url{https://github.com/AndreevP/wvmos}} Then, the cosine similarity \textbf{(cos-sim)} between the generated utterances and the reference speaker embeddings was computed. Furthermore, we compare the distribution of log-F0 values of the generated utterances to the reference VCTK audio samples. For this evaluation, 10 utterances from each of 10 unseen male speakers and 10 unseen female speakers were selected from the VCTK dataset.\footref{foot:embed} 

\section{Results}
\label{section:result}

Table \ref{tab:mos_table_1} shows the results of subjective and objective evaluations. The standard deviations of N-MOS, NR-MOS, and S-MOS scores lie between $1.00$ and $1.24$, while those of WV-MOS and cos-sim scores are approximately $0.3$ and $0.04$ respectively for each model. The best models have been selected (and highlighted) in each row by computing the $95\%$ confidence intervals (CI) on the difference between a pair of models. The difference is statistically significant if the confidence interval lies fully on the positive side of the real axis. 

\subsection{Naturalness and Quality of Utterances}
We found a significant improvement in the naturalness (N-MOS) of speaking utterances generated by GlowTTS-STD and GlowTTS-STDP over the baseline model. This shows that the stochastic duration predictor more accurately learns the speaking style of speakers than the baseline duration predictor. In addition, stochastic pitch prediction does not degrade the naturalness of utterances, and in particular, improves or maintains the naturalness of utterances. For unseen male speakers, we see an improvement of 0.29 MOS points over GlowTTS-STD. 

For longer sentences, the utterances generated by Glow\-TTS-STD and GlowTTS-STDP are rated higher than the baseline in general as measured by NR-MOS. Here, utterances generated by GlowTTS-STDP for female speakers are rated higher than utterances generated by GlowTTS-STD, but there is no significant difference in the ratings for male speakers. In general, short utterances were rated higher than longer utterances. Nevertheless, GlowTTS-STDP still shows significant improvement in reading-style naturalness over the baseline. In addition, the utterances generated by GlowTTS-STD and GlowTTS-STDP have higher overall quality (WV-MOS), however, the ratings are still a distance away from the VCTK utterances with copy synthesis, which were rated higher in general.

\subsection{Diversity of Utterances}
\begin{figure}[h!]
  \centering
    \includegraphics[width=7.0cm,trim=0 10 0 25,clip]{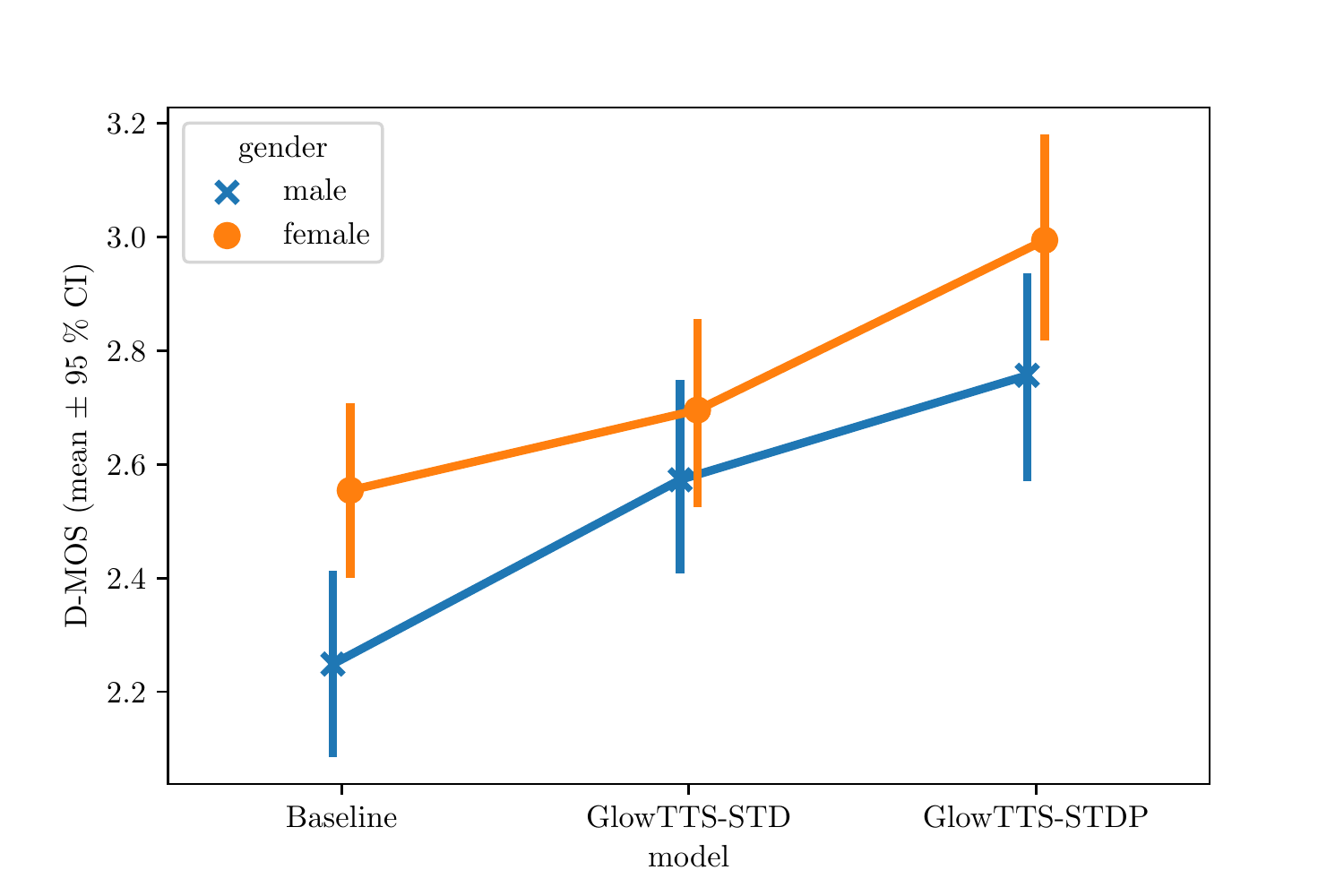}
  \caption{D-MOS of utterances  generated by the evaluated models for both male and female unseen speakers.}
  \label{fig:diversity_distribution}
\end{figure}

In Figure \ref{fig:diversity_distribution}, we see an increasing trend in the diversity of utterances (D-MOS) from the baseline to GlowTTS-STDP, with GlowTTS-STDP being able to generate more diverse utterances than the other models without loss of naturalness. In general, female utterances were rated higher in diversity than male ones.

\begin{figure}[h!]
  \centering
    \includegraphics[width=9.0cm,trim=0 10 0 10,clip]{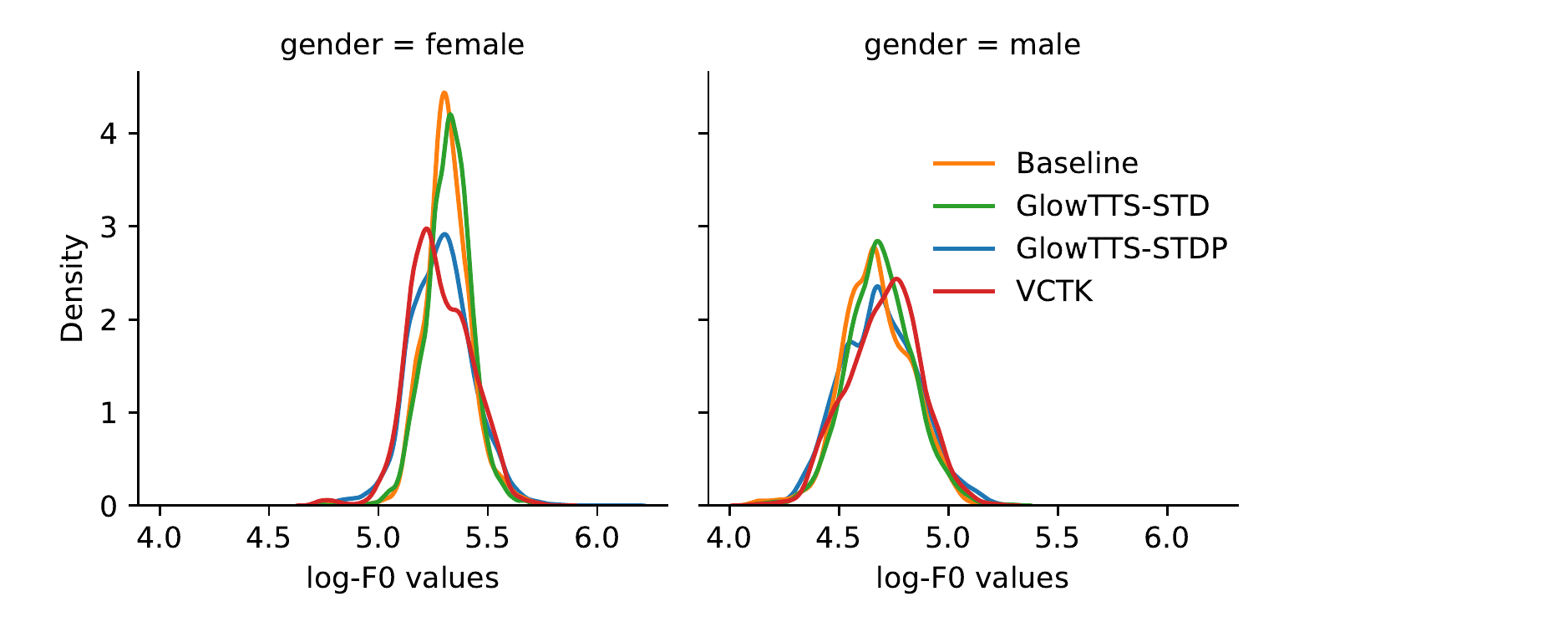}
  \caption{Log-F0 distribution of utterances generated with GlowTTS, GlowTTS-STD, GlowTTS-STDP, and the real VCTK utterances for both male and female speakers. Outliers due to F0 estimation errors have been removed.}
  \label{fig:pitch_distribution}
\end{figure}

Figure \ref{fig:pitch_distribution} shows the distribution of log-F0 values computed for the generated utterances.
Here, we see that the distribution of log-F0 values in GlowTTS-STDP utterances better matches those of real speech for both male and female speakers with more spread-out log-F0 values, while those of the baseline and GlowTTS-STD are narrower, indicating less variability in their predicted log-F0 values.

\subsection{Speaker Similarity} 
The utterances generated by models using a stochastic duration predictor (GlowTTS-STD and GlowTTS-STDP) exhibit significantly higher speaker similarity (S-MOS) than the baseline, especially for female speakers. We believe that the rhythm of speech, controlled by the stochastic duration predictor, significantly impacts the subjective speaker similarity ratings given by volunteers. Similarly, the cos-sim values are higher for GlowTTS-STD and GlowTTS-STDP, however this is not as significant as the S-MOS scores.

\section{Conclusions}
\label{section:conclusion}
In this paper, the naturalness of utterances was successfully improved using a stochastic duration predictor and stochastic pitch predictor. Similarly, we significantly improve the naturalness of long, reading-form utterances. In addition, we increase the diversity of generated utterances by integrating a stochastic pitch predictor into Glow-TTS. We show that the distribution of log-F0 values of our model better matches the distribution of log-F0 values of real utterances. This work is important, as it provides a means to generate more realistic and diverse utterances for downstream tasks e.g., data augmentation for automatic speech recognition and long-form audiobooks. It also enables us to perform procedural manipulation of the rhythm and the pitch of generated utterances, given the prior distribution.

\section{Acknowledgements}
\label{sec:ack}
\ifinterspeechfinal 
Experiments presented in this paper were carried out using the Grid'5000 testbed, supported by a scientific interest group hosted by Inria and including CNRS, RENATER and several Universities as well as other organizations (see \url{https://www.grid5000.fr}).
\else
     This is left blank to meet double-blind requirements.
\fi

\bibliographystyle{IEEEtran}
\bibliography{mybib}

\end{document}